\documentstyle[aps]{revtex} 
\begin{document}
\title{Magnetic ordering in GaAlAs:Mn double well structure}
\author{M. A. Boselli$^1$, I. C. da Cunha Lima$^{1,2}$, and A. Ghazali$^2$}
\address{$^1$ Instituto de F\'{\i}sica, Universidade do Estado do \\
Rio de Janeiro\\
Rua S\~{a}o Francisco Xavier 524, 22.500-013 Rio de Janeiro, R.J., Brazil\\
$^2$ Groupe de Physique des Solides, UMR 7588-CNRS, Universit\'es Paris 7\\
et Paris 6\\
Tour 23, 2 Place Jussieu, F-75 251 Paris Cedex 05, France}
\maketitle
\date{}

\begin{abstract}
The magnetic order in the diluted magnetic semiconductor barrier of double
AlAs/GaAs: Mn quantum well structures is investigated by Monte Carlo
simulations. A confinement adapted RKKY mechanism is implemented for
indirect exchange between Mn ions mediated by holes. It is shown that,
depending on the barrier width and the hole concentration a ferromagnetic or
a spin-glass order can be established.
\end{abstract}


\newpage

In Ga$_{1-x}$Mn$_x$As \cite{van1,van2,oiwa1,oiwa2,matsu,ohno1,ohno2}, a new
prototype of Diluted Magnetic Semiconductors (DMS) \cite{dms1,dms2}, the Mn$%
^{2+}$ cations have the $3d$ shell partially filled with five electrons, in
such a way that they carry a magnetic moment with $S=5/2$. Besides, the Mn
ion binds a hole to satisfy charge neutrality, what is, in itself, a
complicate impurity problem. Two Mn ions occupying nearest neighbors
positions interact {\it via} an anti-ferromagnetic coupling of their
magnetic moments. In the {\it fcc} alloys, these interactions are known to
be frustrated (see, e.g., Ref. \onlinecite{dms1}), establishing the
possibility of settling a spin glass phase at low temperature. However, an
indirect Ruderman-Kittel-Kasuya-Yosida (RKKY) exchange provided by the sp-d
interaction between the Mn$^{2+}$ spins and the spins of the Fermi gas
(unbound holes, or holes in an impurity band) competes with the nearest
neighbors anti-ferromagnetic interaction. At low magnetic ion concentration
the indirect exchange mechanism may become the dominant interaction, leading
either to a spin-glass phase or a ferromagnetic order.

Recently some groups \cite{van1,van2,oiwa1,oiwa2,matsu,ohno1,ohno2}
succeeded in producing homogeneous samples of Ga$_{1-x}$Mn$_{x}$As alloys
with $x$ up to $7\%$ avoiding the formation of MnAs clusters by using low
temperature ($200-300^{o}$ C) MBE techniques. Besides its practical
importance, this kind of DMS introduces an interesting problem from the
physical point of view: Mn in the alloy is a strong $p$ dopant, the free
hole concentration reaching even $p=10^{20-21}cm^{-3}$ \cite
{van1,van2,oiwa1,oiwa2,matsu,ohno1,ohno2}. At small Mn concentrations, the
alloy is a paramagnet and an insulator. As $x$ increases it becomes
ferromagnetic, going through a non-metal/metal transition for higher
concentrations ($x\approx 0.03$), and keeping its ferromagnetic phase. For $x
$ above $7\%$, the alloy becomes a ferromagnetic insulator. In the metallic
phase, the ferromagnetic transition is observed in the range of $30-100$K,
depending on the value of $x$. The ferromagnetic order in the metallic phase
is understood as resulting from the indirect exchange between the Mn ions
mediated by the hole gas. In quantum wells, it seems to exist a threshold
for the width of the magnetic layer in order for a ferromagnetic phase to
appear \cite{haya}.

In this work we perform Monte Carlo calculations to study the magnetic order
resulting from the indirect exchange between magnetic moments in a
particular symmetric double quantum well DMS structure formed by two GaAs
wells of width $L$ separated by a Ga$_{0.65}$Al$_{0.35}$As : Mn DMS barrier
width $d$. In that DMS the Mn$^{2+}$ ion concentration is taken as $x=0.05$.
The Mn$^{2+}$ ions substitute the cations elements, each of them providing a
hole. In addition, the DMS is assumed to be in a metallic phase, but the
density of free carriers (holes) is only a fraction $r$ of the magnetic ion
concentration, what is in agreement with experimental data of Ref.%
\onlinecite{ohno1}. A confinement-adapted RKKY \cite{bose} formalism is used
to obtain the indirect exchange for the double quantum well structure: 
\begin{equation}
H_{ex}=-\sum_{i<j}J_{ij}\vec S_i\cdot \vec S_j,  \label{excham}
\end{equation}
\begin{equation}
J_{ij}=\left( \frac{I}{2A}\right) ^{2}\sum_{n,n^{\prime }}\sum_{\vec{q}}2 %
\mbox{\rm Re} \left[ \phi _{n}^{\ast }(z_{i})\phi _{n^{\prime }}(z_{i})\phi
_{n^{\prime }}^{\ast }(z_{j})\phi _{n}(z_{j})e^{-i\vec{q}.(\vec{R}_{i}-\vec{R%
}_{j})}\right] \chi ^{n,n^{\prime }}(\vec{q}).  \label{generj}
\end{equation}
with $\phi_n(z)$ representing the eigenfunctions of the potential well, $I$
the sp-d interaction \cite{tward,okabayashi,bhattacharjee}, and $A$ the
normalization area for the otherwise free motion in the ($x,y$) plane. The
coordinates ($\vec R_i,z_i$) describe the position of the impurity $i$ in
the plane ($x,y$), and in the growth direction inside the barrier. $\chi
^{n,n^{\prime }}(\vec q)$ is the equivalent to the Lindhard function: 
\begin{equation}
\chi ^{n,n^{\prime }}(\vec q)=\sum_{\vec k}\frac{\theta (E_F-\epsilon _{n,%
\vec k})-\theta (E_F-\epsilon _{n^{\prime },\vec k+\vec q})}{\epsilon
_{n^{\prime },\vec k+\vec q}-\epsilon _{n,\vec k}}.  \label{modlin}
\end{equation}

It is worthwhile to mention that the hole system is expected to show
a spin polarization in a DMS magnetically ordered phase.
This effect, which is important in spin resonant tunneling experiments,
does not result into a major change in the order of the Mn$^{++}$ impurities.

The intra-subband contribution of an occupied subband $n$ to the exchange
reads: 
\begin{eqnarray}
J_{ij}^{(n)} &=&-\left( \frac{I}{2}\right) ^{2}\frac{m_{t}^{\ast }}{\pi
\hbar ^{2}}k_{F}^{(n)2}\mid \phi _{n}(z_{i})\mid ^{2}\mid \phi
_{n}(z_{j})\mid ^{2}\times  \nonumber \\
&&[J_{0}(k_{F}^{(n)}R_{ij})N_{0}(k_{F}^{(n)}R_{ij})+J_{1}(k_{F}^{(n)}R_{ij})N_{1}(k_{F}^{(n)}R_{ij})].
\label{jintra}
\end{eqnarray}
where $m_t^{\ast}$ is the transversal effective mass, and $k_F^{(n)}$ is the 
$n$-th subband Fermi wave vector.

The contribution of the inter-subband terms cannot be expressed easily in a
closed form. Starting over from Eq. (\ref{generj}) we arrive to: 
\begin{equation}
J_{ij}^{(n,n^{\prime })}=\left( \frac{I}{2}\right) ^{2}\frac{1}{\pi }%
\mbox{\rm Re}\left[ \phi _{n^{\prime }}^{\ast }(z_{i})\phi _{n}(z_{i})\phi
_{n}^{\ast }(z_{j})\phi _{n^{\prime }}(z_{j})\right] \int_{0}^{\infty
}dqqF_{n,n^{\prime }}(q)J_{0}(qR_{ij}),  \label{stj}
\end{equation}
where we used 
\begin{equation}
F_{n,n^{\prime }}(q)=\frac{4m_{t}^{\ast }}{(2\pi \hbar)^{2}} \int d^2k \frac{%
q^{2}+\Delta _{n^{\prime },n}}{(q^{2}+\Delta _{n^{\prime },n})^{2}-(2\vec{k}%
\cdot \vec{q})^{2}}\theta (E_F-\epsilon _{n,\vec{k}}),  \label{deff}
\end{equation}
and $\Delta _{n^{\prime },n}=2m_{t}^{\ast }\cdot (E_{n^{\prime
}}-E_{n})/\hbar ^{2}$. The integral in Eq. (\ref{deff}) is, then,
straightforward: 
\begin{equation}
F_{n,n^{\prime }}(q)=\frac{m_{t}^{\ast }}{2\pi \hbar ^{2}}(1+\frac{\Delta
_{n^{\prime },n}}{q^{2}})[1-\sqrt{1-(\frac{2k_{F}^{(n)}q}{q^{2}+\Delta
_{n^{\prime },n}})^{2}}\theta (q^{2}+\Delta _{n^{\prime
},n}-2qk_{F}^{(n)})]\theta (E_F-\epsilon _{n}).
\end{equation}

It is well known \cite{bose,dietl} that, in the mean field approximation, a
DMS quantum well with infinite barriers has no inter-subband contribution to
the Curie-Weiss temperature. For finite barriers and low carrier
concentrations, these contributions are also small. That is not the case,
however, when a DMS layer is inserted in the middle of a non-magnetic
quantum well \cite{bose}, as in the present structure.

A Monte Carlo (MC) simulation was performed to investigate the magnetic
order in the DMS layer. The spin sites in that layer belong to the cation 
{\it fcc} sublattice. They are distributed randomly and are occupied by Mn$%
^{2+}$ ions, with a concentration $x$. The calculation is performed in a
finite box whose axes are parallel to [100] directions, of dimensions $%
L_{x}=L_{y}$, and $L_{z}=N\ a/2$, where $a$ is the lattice parameter of
GaAs, and $N$ the number of DMS monolayers (ML) in the barrier. Periodic
boundary conditions are imposed in the $(x,y)$ plane. Lateral dimensions are
adjusted in such a way that the total number $N_{s}$ of spins is about 4400,
for all $L_{z}$. Their initial orientations are randomly assigned. The
energy of the system due to RKKY interaction described by the above $J_{ij}$%
's is calculated, and the equilibrium state for a given temperature is
sought by changing the individual spin orientation according to the
Metropolis algorithm \cite{diep}. A slow cooling stepwise process is
accomplished making sure that thermal equilibrium is reached at every
temperature. The resulting spin configuration is taken as the starting
configuration for the next step with a lower temperature. For every
temperature, the average magnetization $<M>$ and the Edwards-Anderson order
parameter $q$ are calculated. The latter is defined as 
\begin{equation}
q=(1/N_{s})\sum_{i}\sqrt{(1/T)%
\sum_{t}[S_{x,i}(t)^{2}+S_{y,i}(t)^{2}+S_{z,i}(t)^{2}]}  \label{EAq}
\end{equation}
where the expression under the square root is the MC time average.

We present results for the eight samples described in Table 1. $L$ was
chosen typically 6 nm. The DMS barrier width was varied from 4 ML to 18 ML
and the density fraction $r$ was taken equal to 0.1 and 0.25. In Fig. 1 the
normalized magnetization $<M>$ is plotted {\it versus} temperature. Samples
\# 3, \# 5 and \# 8 show a ferromagnetic order at $T_{c}\approx $35, 55 and
15 K, respectively. The complete saturation at T = 0 K is not achieved,
presumably due to boundary effects in the z-direction. It is striking that
the magnetization curves in those samples are far from the canonical
Brillouin function, as already experimentally remarked in Ref. %
\onlinecite{van1}, being instead very similar to the ones obtained by these
authors. On the other hand, samples \# 2 and \# 7 are still paramagnetic at
T $\approx$ 1 K. The Edwards-Anderson order parameter $q$ given by Eq. \ref
{EAq} is shown in Fig. 2 as a function of temperature. The ferromagnetic
samples have nearly the same $q(T)$ as $<M>(T)$, what is not surprising.
However, samples \# 1, \# 4 and \# 6 show large $q$-values at low
temperature, indicating a spin-glass like magnetic order. In particular,
samples \# 4 and \# 6 show a non negligible magnetization at low
temperatures. This points to an occurrence of a canted spin phase.

The results seem to indicate that two relevant parameters compete in
establishing the magnetic order in the DMS layer. On one hand, a
ferromagnetic order is expected to settle as the layer width is increased,
in accordance with what observed in Ref. \onlinecite{haya}. On the other
hand, depending on the hole density, and because of the oscillatory nature
of RKKY interaction, in addition to the ferromagnetic couplings,
antiferromagnetic couplings can be switched on. This is the ingredient which
together with disorder are the origin of the spin-glass phase. This is
illustrated in Fig. 3 where the RKKY interactions $J_{ij}$ are plotted {\it %
versus} the in-plane distance $R_{ij}$ for samples \# 5, \# 6 and \# 8. One
sees that for ferromagnetic samples (\# 5 and \# 8) $J_{ij}$ is essentially
positive, while for sample \# 6, a non negligible negative part is present,
especially when considering the occurrence of pair couplings with large $%
R_{ij}$'s. Therefore it is not surprising that a (canted) spin-glass phase
sets in.

In conclusion, we have shown that depending on the DMS barrier width of a
double QW structure and the hole concentration, different magnetic phases
can be obtained. For applications, favorable spin configurations can thus be
designed. In particular, the possibility of having a ferromagnetic order in
the DMS barrier of a double QW structure is important for spin tunneling and
resonant spin tunneling in nanostructures \cite{ohno1,bruno}.

This work was partially supported by CAPES, FAPERJ and CENAPAD-SP
(UNICAMP-FINEP/MCT) in Brazil, and by the
PAST grant from Minist\`{e}re de l'\'{E}ducation Nationale, de
l'Enseignement Sup\'{e}rieure et de la Recherche (France).

\newpage 
\begin{table}[tbp]
\caption{Samples characteristics: The width of each one of the two GaAs well
in the structure is $L=6 nm$; N is the number of DMS barrier monolayers
(ML), $r$ is the ratio of hole density to Mn density; $T_c(K)$ is the
magnetic transition temperature for F: ferromagnetic, P: paramagnetic, SG:
spin-glass phases. The sign $(*)$ indicates a possible canted phase. The
calculation is performed with the value of $N_0 \beta = -1.2$ eV
according to ref.{\protect \onlinecite{okabayashi}}. }
\label{tab01}
\begin{tabular}{|c|cccc|}
Sample & $N(ML)$ & $r$ & Phase & $T_c(K)$ \\ \hline
\# 1 & 4 & 0.1 & SG & 2 \\ 
\# 2 & 12 & 0.1 & P & $\leq$ 1 \\ 
\# 3 & 18 & 0.1 & F & 35 \\ 
\# 4 & 4 & 0.25 & SG$^*$ & 15 \\ 
\# 5 & 18 & 0.25 & F & 55 \\ 
\# 6 & 12 & 0.25 & SG$^*$ & 30 \\ 
\# 7 & 9 & 0.1 & P & $\leq$ 1 \\ 
\# 8 & 9 & 0.25 & F & 15 
\end{tabular}
\end{table}
\newpage

\begin{figure}[tbp]
\caption{Normalized magnetization {\it vs} temperature for samples indicated
in Table 1.}
\label{fig1a}
\end{figure}

\begin{figure}[tbp]
\caption{Edwards-Anderson order parameter $q$ {\it vs} temperature for
samples indicated in Table 1.}
\label{fig1b}
\end{figure}

\begin{figure}[tbp]
\caption{RKKY exchange interaction $J_{ij}$ {\it vs} $R_{ij}$ for samples \#
5, \# 6 and \# 8.}
\label{fig2}
\end{figure}

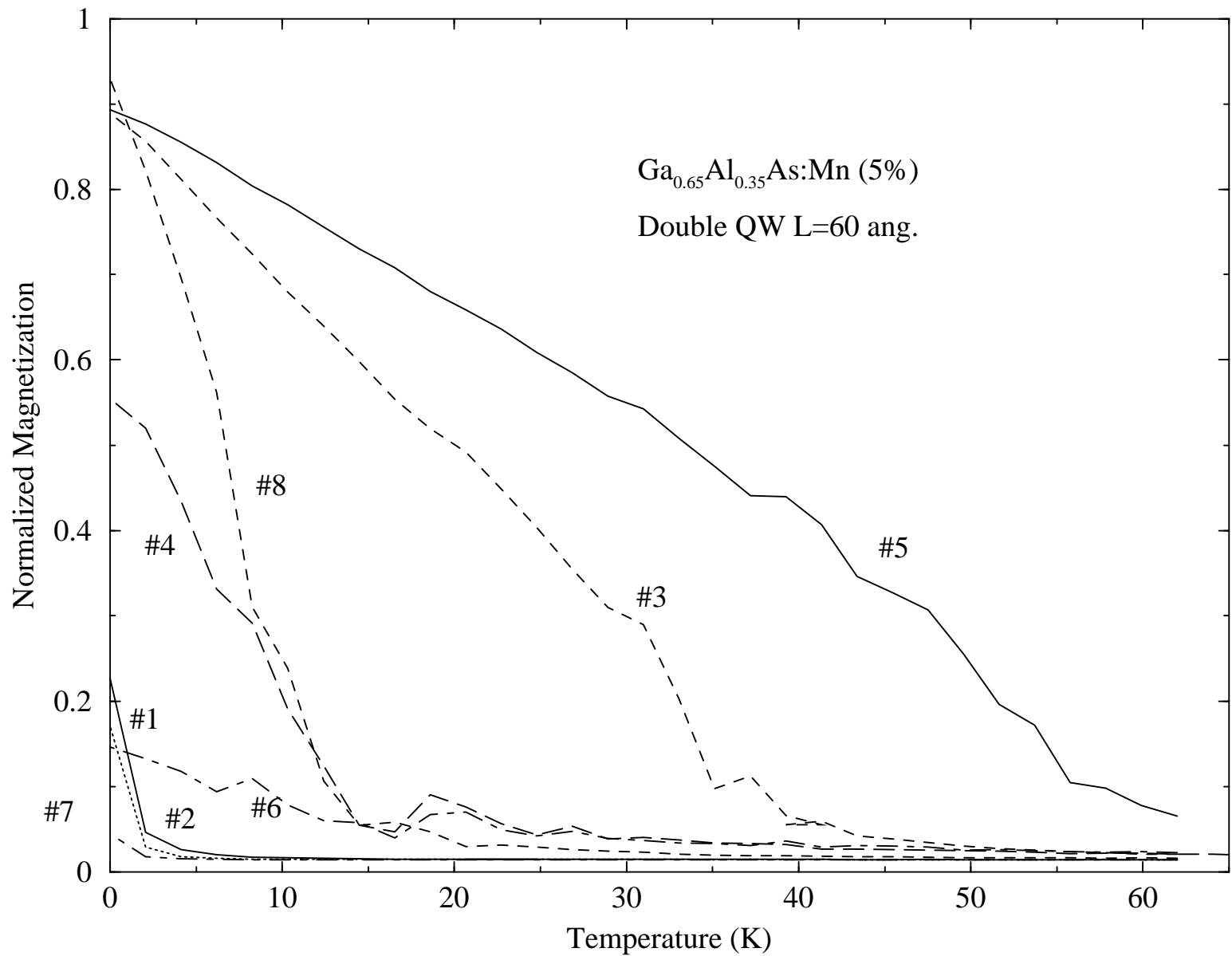
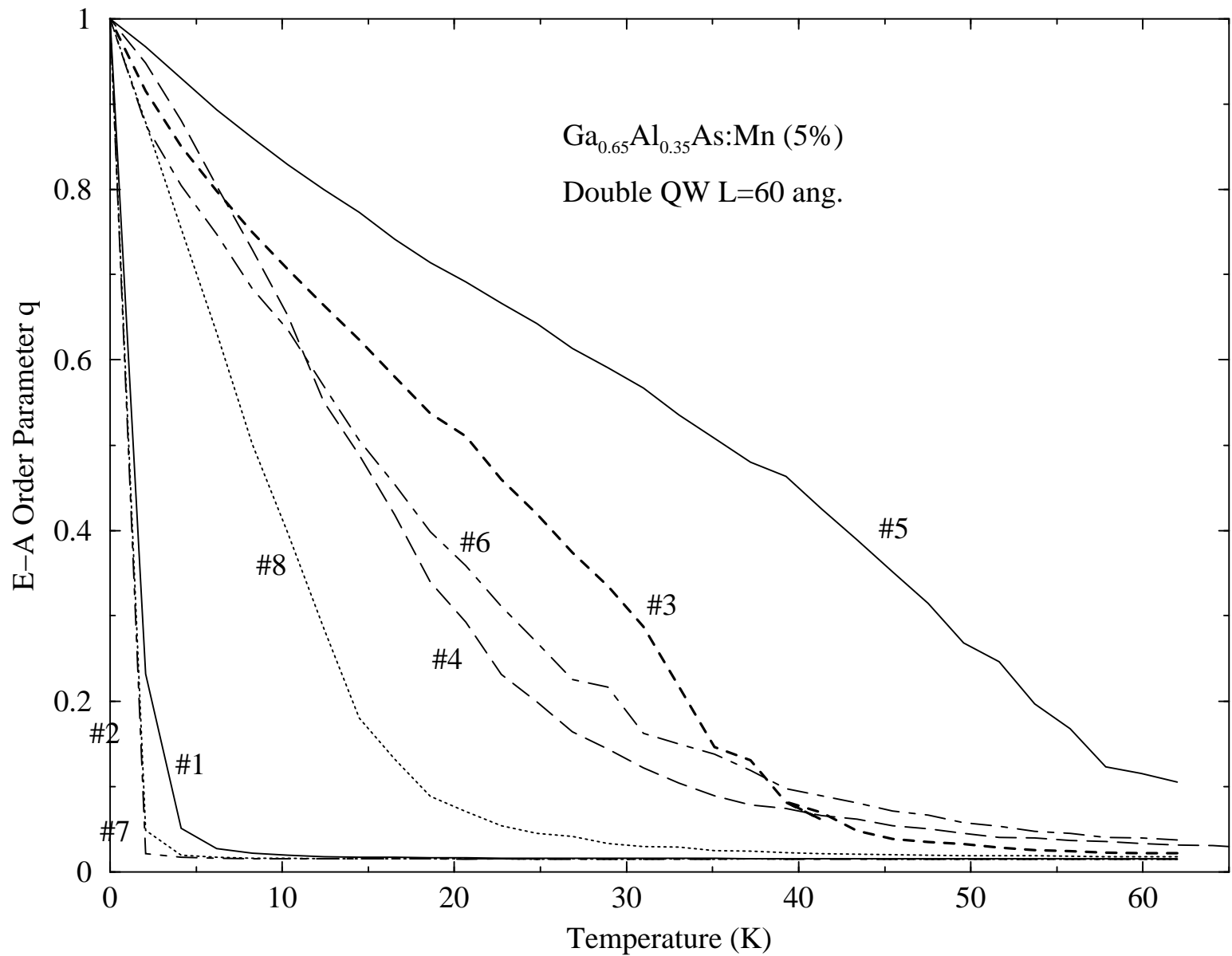
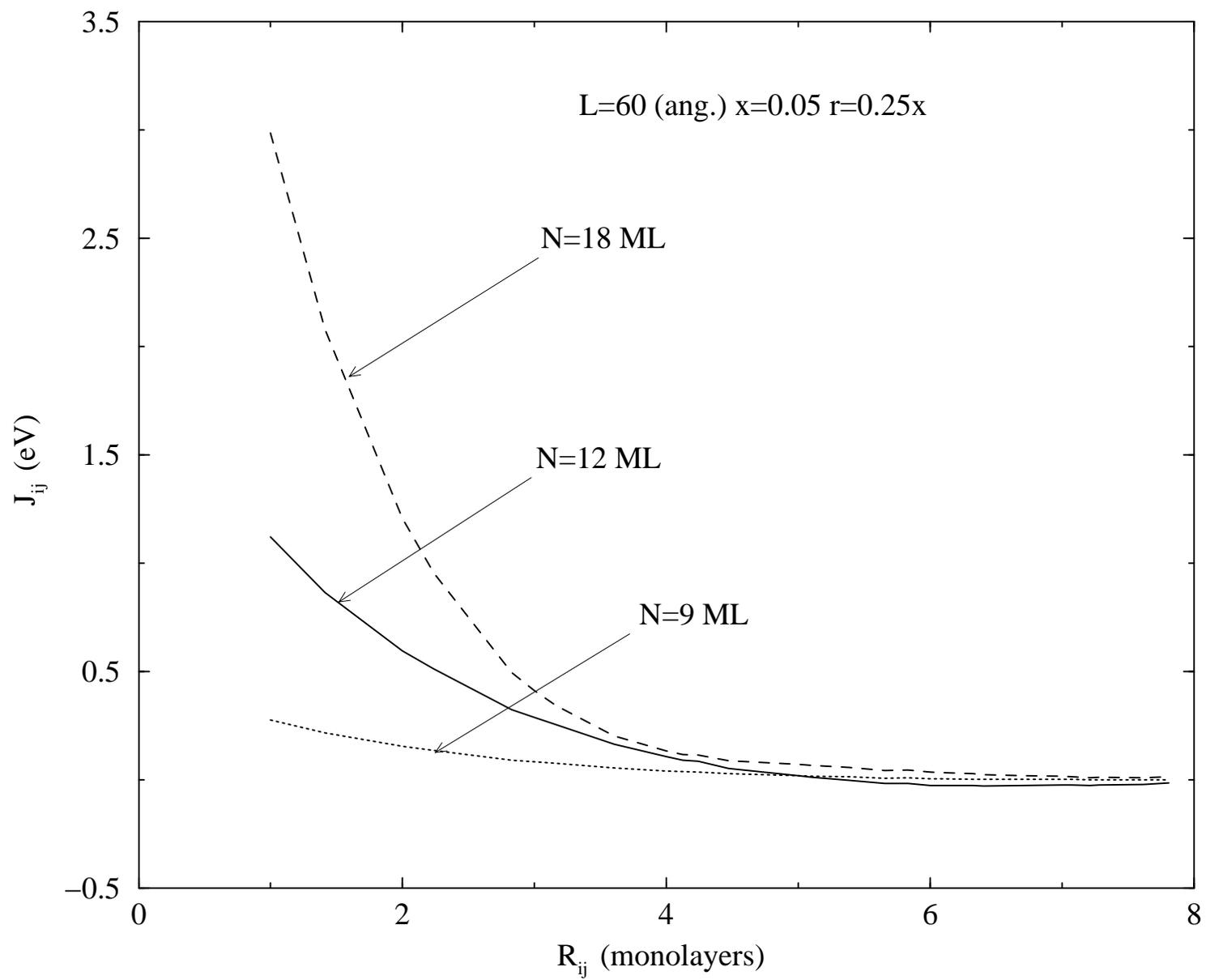
\end{document}